# Estimation of the shear strength of a coarse soil with high fines content by parallel gradation method

*N'guessan Moïse* Kouakou[1,2*], *Olivier* Cuisinier[1], *Farimah* Masrouri[1], *Emmanuel* Lavallée[2], and *Tangi* Le Borgne[2]

[1]LEMTA (UMR7563, Université de Lorraine – CNRS), 54500 Vandœuvre-lès-Nancy, France
[2]Bouygues Travaux Publics, 78000 Guyancourt, France

**Abstract.** The determination of the mechanical properties of soils containing particles larger than the allowable size of standard laboratory equipments is complex. It is indeed necessary to remove the coarsest fraction to carry out the tests. This scalping poses a problem of reliability of the results at the scale of the structure. Parallel gradation is the method commonly used for estimating the shear strength of heterogeneous granular soils from tests on their finer fraction. However, the effect of high fines content on the estimation of shear strength by this method is not well understood. The results of this study showed that the parallel gradation method could predict the friction angle of the initial soil with high fines content when the modelled soil had a similar skeleton as the initial soil. However, the cohesion of the initial soil was overestimated.

## 1 Introduction

Heterogeneous granular soils with matrix are composed of variable grains size (from a few microns to several tens of centimetres). Soil types in this category can be of natural deposit (alluvium, scree, moraine, etc.) or made by humans (rockfill). These soils are often used in the construction of dikes and embankments. Their applications in geotechnical projects require the determination of their mechanical characteristics in laboratory, which is a challenge due to the presence of coarse grains. Indeed, the maximum grain size of these soils is larger than the maximum size allowed by common laboratory shear devices. A solution is to use large devices. For example, a direct shear box of 1 m$^3$ was used to determine the failure parameters of rock aggregates [1] and a triaxial cell of 1 m in diameter and 1.5 m in height was used for studying rockfills [2]. However, these devices, which allow to test soils with a grain size up to 160 mm, are few, complex to implement and unsuitable for materials with larger grain size.

Therefore, some authors have developed methods to estimate with good accuracy the mechanical parameters of these soils from tests on their fraction compatible with standard laboratory shear equipments. Three modelling techniques have been proposed in the literature: substitution, scalping, and parallel gradation.

Among these approaches, parallel gradation was the most used because it preserves the grain size distribution curve of the initial soil. It consists in scalping the oversize grains and reconstituting a material with a grain size distribution parallel to the initial soil. When this method was used, the modelled material had a greater friction angle than the original material [3]. However, the authors showed that the friction angles of the two materials are identical for the same grain breakage ratio. Based on these results, an analytical method was proposed to take into account the effect of grain breakage during shearing and thus have a good estimation of the shear strength of the initial material [4]. Furthermore, some authors concluded that the method led to an underestimation of the initial soil shear strength when the breakage ratio was low and an overestimation when the breakage ratio was high [5]. In addition, most of the studies were carried out on materials with low fines content. Therefore, some questions remain about the effect of high fines content in the modelled material. For soils with high fines content, some authors proposed to attribute the fines content of the initial soil to the modelled soil. But a study showed that this underestimates the shear strength of the initial soil [6]. Therefore, even if promising results were obtained by the parallel gradation method on coarse granular soils with matrix, the effect of higher fines content in the modelled soil remains unknown.

This study focused on the estimation of the shear strength of heterogeneous granular soils with a high fines content (> 10%) by using the parallel gradation method. The main question was whether the parallel gradation method could give a good estimation of shear strength despite a high fines content in the initial material. For this purpose, the failure criteria of an initial soil with a high percentage of fines and two modelled soils obtained by parallel gradation method were determined and compared.

[*] Corresponding author: n-guessan-moise.kouakou@univ-lorraine.fr

## 2 Material and methods

### 2.1. Soils characteristics

A natural gravel with particles sizes up to 120 mm was chosen as the support soil for the study. It contains 4.3% of fines (particles smaller than 0.080 mm). Since the dimensions of the available equipments (see section 2.2) require a maximum grain size of 30 mm, particles larger than 30 mm were removed. The 0-30 mm soil with grading curve parallel to the initial soil was chosen as the prototype soil for shear tests. Its maximum dry bulk density determined from a Proctor test is 21.1 kN/m$^3$ for an optimal water content of 8.3%. From the prototype soil, two modelled soils with parallel grading curve were made (0-15 and 0-5 mm). The grading curves of each material are shown in Figure 1. The fines content of 0-30, 0-15 and 0-5 mm specimens are 15; 19.8 and 22.5% respectively.

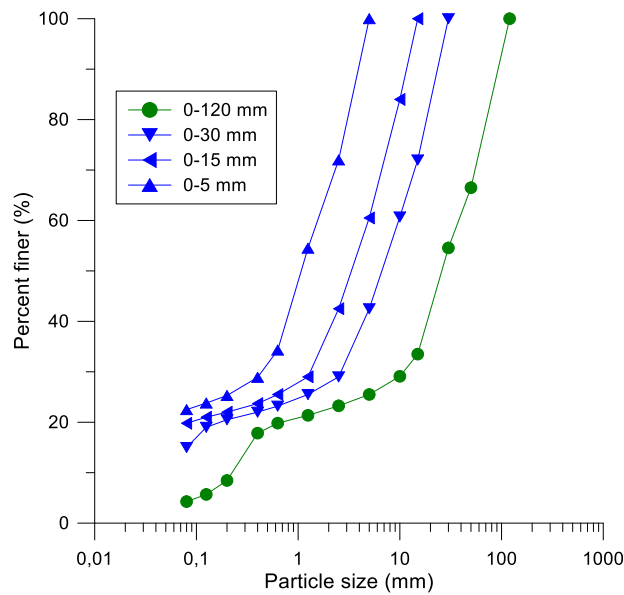

**Fig. 1** : Grain size distributions of studied materials

### 2.2 Direct shear apparatuses

Three direct shear boxes were used: a small box of 60 x 60 x 45 mm$^3$, a medium box of 150 x 150 x 180 mm$^3$ and a large box of 300 x 300 x 180 mm$^3$. The small box was used for 0-5 mm specimens, the medium box for 0-15 mm specimens and the large box for 0-30 mm specimens.
The small box is a classic direct shear box for soil mechanics.
The large direct shear box is composed of two half boxes: the lower half box is fixed while the upper half box is pushed by a motor at constant velocity. The cross-section of the large box can be reduced to 150 x 150 mm$^2$ using adapters to get the medium box.
A system of frame and hydraulic piston with a maximum capacity of 100 kN applied a constant normal load during the test. Displacement sensors were installed to record the horizontal and vertical movements of the specimen. These sensors had a range of 50 mm with an accuracy of 0.2%. The shear force was measured by a load sensor with a capacity of 100 kN. The normal load was measured by a load sensor placed between the frame and the upper plate. An inclination sensor installed on the upper plate recorded its inclination movements during the test. All the sensors were connected to a central acquisition unit which transmits the data to a software for recording and visualization.

### 2.3 Experimental strategy

The purpose of the tests was to know if the parallel gradation method could be used to predict the shear strength of coarse soil with high fines content. For this purpose, the modelled materials 0-5, 0-15 and 0-30 mm with grain size distribution parallel to the initial material were tested at the same dry density of 2 Mg/m$^3$. The same density was chosen for all specimens for a better comparison of their shear strengths [4]. The failure criteria of these specimens were determined from consolidated drained tests and compared.

### 2.4 Samples preparation

In order to reconstitute the desired grain size for the specimens, the particles size up to 30 mm of the initial sample was separated into different fractions by sieving: < 0.08; 0.08-0.4; 0.4-0.63; 0.63-1.25; 1.25-2.5; 2.5-5; 5-10; 10-15 and 15-30 mm. The different fractions were mixed according to the proportions given by the grading curve for the preparation of the specimens. All the specimens were prepared at the optimal water content (8.3%) and kept in airtight bags for at least 24 hours to ensure good homogeneity of the water in the samples. The specimens were then statically compacted until the desired density was reached: in three layers inside the medium or large shear box and in one layer inside the small box. The specimens were saturated by filling the shear apparatus tank with demineralised water and allowing it to flow through the sample for at least three hours. After saturation, the specimens were consolidated under the desired normal stress and then sheared at a constant rate of 0.05 mm/min.

## 3 Results

### 3.1. Parallel gradation stress-strain curves

The stress-strain curves and the volumetric strain responses of the three specimens tested under different normal stresses were presented in Figures 2 to 4. The stress-strain curves showed a distinct peak and residual values for 0-5 mm specimen. In the case of 0-15 and 0-30 mm specimens, there was no apparent peak while all specimens were tested at the same dry density. The absence of a peak for 0-15 and 0-30 mm specimens could be explained by the large dimensions of the shear box used. Indeed, some authors have found that as the shear box dimensions increase, the difference between peak and residual strength decreases until it disappears completely [7, 8]. Also, the maximum shear values obtained for the 0-15 mm specimen were substantially equal to the peak values for the 0-5 mm specimen, except under the stress of 100 kPa. As for the 0-30 mm specimen, its maximum shear strengths were between the peak and residual values of the 0-5 mm specimen.

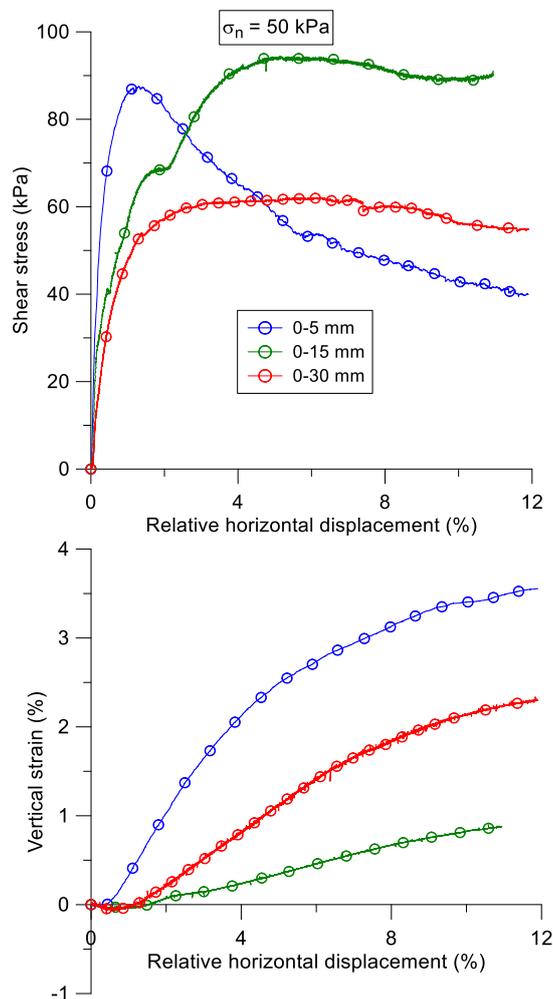

**Fig. 2 :** Stress-strain curves of studied samples under a normal stress of 50 kPa

The volumetric curves started with a contractance phase followed by a dilation phase characteristic of the dense state of the specimens in all boxes. The contraction of the 0-5 mm material was very low regardless the normal stress with a maximum value of 0.05% while it occurred over a longer deformation interval with a higher value for 0-15 and 0-30 mm specimens. However, the dilation had a higher value for 0-5 mm material compared to 0-30 mm material. The specimen 0-15 mm had lowest dilation values whereas they were expected between those of 0-5 and 0-30 mm specimens. This could be explained by the higher height/width ratio of the box for this specimen. Indeed, 0-5 and 0-30 mm specimens had a height/width ratio of about 0.6 while 0-15 mm specimen had a ratio of 1.1.

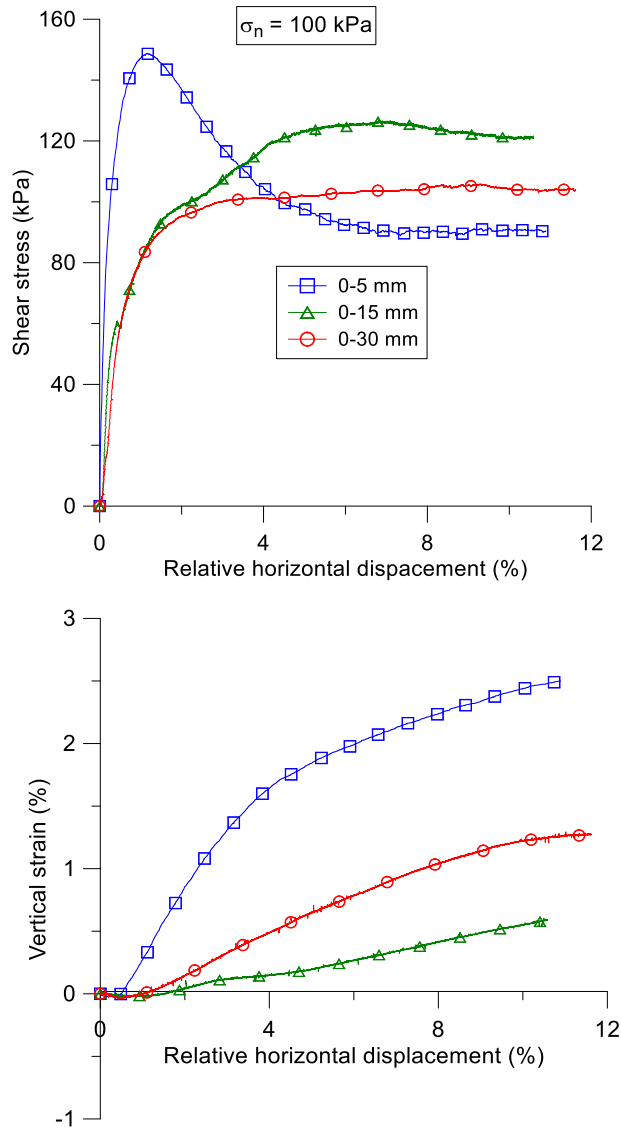

**Fig. 3 :** Stress-strain curves of studied samples under a normal stress of 100 kPa

### 3.2. Failure criteria of studied materials

The failure criteria for parallel gradation specimens according to the Mohr-Coulomb model are shown in Figure 5. The three specimens had a same maximum friction angle. The approach therefore gave a good prediction of the friction angle of coarse soils with high fines content in the range of applied stresses. The high fines content in the studied soil had no effect on the estimation of the friction angle by the method. However, the cohesion increased when increasing fines content.

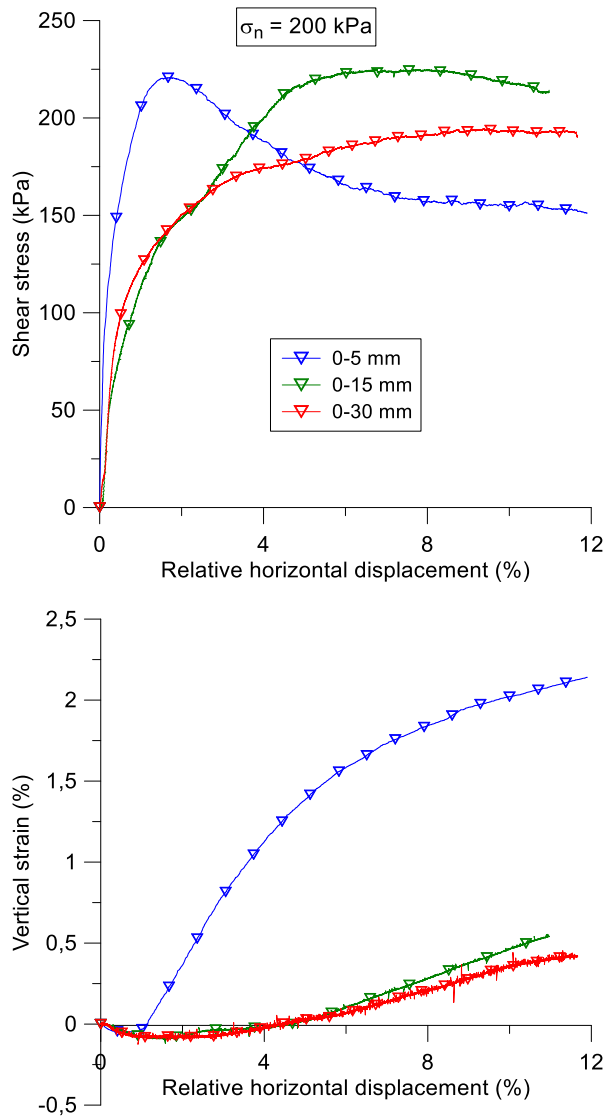

**Fig. 4 :** Stress-strain curves of studied samples under a normal stress of 200 kPa

A result in literature showed that from a certain percentage of fines in the modelled material, the parallel gradation method could no longer predict the angle of friction of the initial material [9]. Indeed, the authors tested two materials of parallel grain size: a 0-25.4 mm specimen with 17% fines and a 0-4.75 mm specimen with 33% fines. They found a higher cohesion and a lower friction angle for the 0-4.75 mm specimen. This difference is probably related to the difference in the skeleton of the two specimens. The 0-25.4 mm specimen had a skeleton imposed by the granular elements of the soil while the 0-4.75 mm specimen skeleton was partially controlled by fines. The parallel gradation method would therefore give a good estimation of the friction angle for a skeleton of the modelled soil identical to that of the initial soil, i.e. a fines content less than 20 to 35% [10, 11].
Considering the obtained results and the fact that the tested materials had a grain size distribution parallel to the 0-120 mm sample, an estimation of the failure parameters of the 0-120 mm soil could be proposed. Thus, a friction angle of 41° and zero cohesion could be attributed to the 0-120 mm soil because it contains low fines content (4.3%).

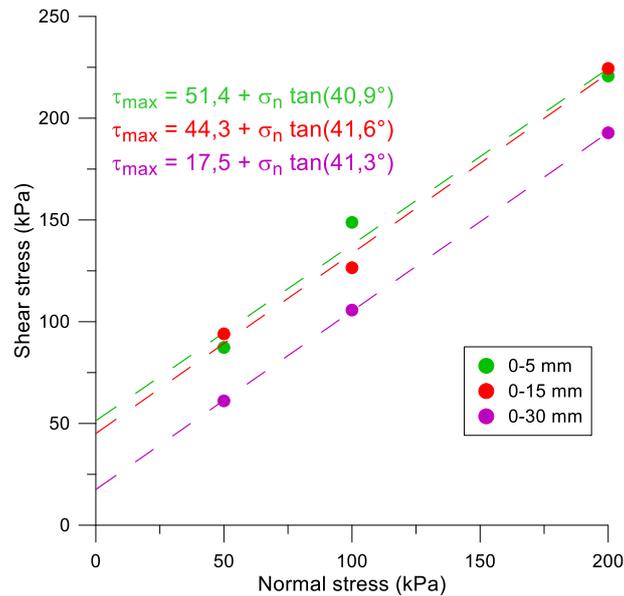

**Fig. 5 :** Shear strength envelopes for 0-5, 0-15 and 0-30 mm

## 4 Conclusions

The aim of this study was to determine the effect of significant fines content in the initial material on the estimation of the shear strength of heterogeneous granular soil by the parallel gradation method. The main conclusion of this study is:
The parallel gradation method gave a good estimation of the maximum friction angle of heterogeneous granular soils with high fines content when the modelled soil had a similar skeleton as the initial soil. However, the significant increase of fines content in the modelled soil led to an overestimation of the initial soil cohesion.
Additional tests are needed to fully understand the evolution of soil skeleton when increasing fines content in the modelled soil.